\documentclass[10pt, conference]{IEEEtran}
\IEEEoverridecommandlockouts

\usepackage{cite}
\usepackage{amsmath,amssymb,amsfonts}
\usepackage{algorithm}
\usepackage{algpseudocode} 
\usepackage{graphicx}
\usepackage{textcomp}
\usepackage{comment}
\usepackage{todonotes}
\usepackage{booktabs}
\usepackage[caption=false,font=footnotesize]{subfig}
\usepackage{xcolor}
\usepackage[hidelinks, bookmarks=false]{hyperref} 
\usepackage{mathrsfs}
\usepackage{array}
\def\BibTeX{{\rm B\kern-.05em{\sc i\kern-.025em b}\kern-.08em
    T\kern-.1667em\lower.7ex\hbox{E}\kern-.125emX}}

\usepackage{letltxmacro}
\usepackage{xparse}
\usepackage{multirow}
\usepackage{amsthm}
\usepackage{pgfplots}
\usepackage{tikz}
\usepackage{mathrsfs}
\usepackage{thmtools}

\usetikzlibrary{arrows, shapes}
\def\equationautorefname~#1\null{%
	Eq.~(#1)\null
}

\definecolor{lgray}{gray}{0.7}
\definecolor{llgray}{gray}{0.83}
\definecolor{lllgray}{gray}{0.92}

\newcommand{\redcomment}[1]{\textcolor{red}{#1}}
\definecolor{greedyColor}{HTML}{404E5C}
\definecolor{bzipColor}{HTML}{35CE8D}
\definecolor{lz4Color}{HTML}{B7C3F3}
\definecolor{zstdColor}{HTML}{DD7596}
\definecolor{lossless}{HTML}{CA895F}

\newcolumntype{A}{>{\centering}m{2cm}}
\newcommand{\w}[1]{#1}
\newcommand{\ws}[1]{#1}
\newcommand{\techniquename}{\emph{Zeal}}
\newcommand{\Ebar}{\bar{\mathrm{H}}}
\newcommand{\Abar}{\bar{\mathrm{A}}}
\newcommand{\point}{x}
\newcommand{\pointstarmin}{\point^*_{\min}}
\newcommand{\pointstarmax}{\point^*_{\max}}
\newcommand{\expectedvalue}{\mathbb{E}}
\newcommand{\C}{\mathrm{C}}
\newcommand{\elle}{\mathrm{L}}
\newcommand{\R}{\mathrm{R}}
\newcommand{\E}{\mathrm{h}}
\newcommand{\variance}{\mathrm{VAR}}
\newcommand{\pcumul}{p_C}
\newcommand{\ulp}{\mathrm{ULP}}
\newcommand{\mantissalength}{L}
\newcommand{\angularcoeff}{a}
\newcommand{\pdf}{\mathrm{PDF}}

\newcommand{\invcdf}{\mathrm{CDF}^{-1}}
\newcommand{\minexpvul}{{E_{U}^{\mathrm{vul}}}}
\newcommand{\minexpencl}{{E_{U}^{\mathrm{enc}}}}
\newcommand{\techperturb}{f^*}
\newcommand{\techaverage}{\mathrm{AVG^*}}
\newcommand{\Deltaavg}{\Delta_{\mathrm{AVG}}}
\newcommand{\deltaavg}{\delta_{\mathrm{AVG}}}
\newcommand{\deltaexpone}{\delta_{\mathbb{E}\left(\point_1\right)}}
\newcommand{\deltaexpi}{\delta_{\mathbb{E}\left(\point_i\right)}}

\newcommand{\sumofds}{\mathrm{S}_{\mathrm{DS}} }

\declaretheorem[name=Theorem, numberwithin=section,
refname={Thm.,Thms.},
Refname={Theorem,Theorems}]{thm}
\declaretheorem[name=Definition,numberwithin=section,
refname={Def,Defs.},
Refname={Definition,Definitions}]{defn}
\declaretheorem[name=Corollary,numberwithin=thm,
refname={Coroll.,Corolls.},
Refname={Corollary,Corollaries}]{corol}

\declaretheorem[name = Remark, style=remark, 
refname={Remark, Remarks},
Refname={Remark,Remarks}]{remark}

\begin{document}
\bstctlcite{IEEEexample:BSTcontrol}
\title{Zip to Zip-it: Compression to Achieve Local Differential Privacy\\
\thanks{This work was supported by the IoTalentum Project within the Framework of Marie Skłodowska-Curie Actions Innovative Training Networks (ITN)-European Training Networks (ETN), which is funded by the European Union Horizon 2020 Research and Innovation Program under Grant 953442.}
}

\author{
	\IEEEauthorblockN{Francesco Taurone, Daniel E. Lucani and Qi Zhang}
	\IEEEauthorblockA{
		DIGIT, Department of Electrical and Computer Engineering, Aarhus University\\
		\{francesco.taurone, daniel.lucani, qz\}@ece.au.dk
	}
}

\maketitle

\begin{abstract}
Local differential privacy techniques for numerical data typically transform a dataset to ensure a bound on the likelihood that, given a query, a malicious user could infer information on the original samples. Queries are often solely based on users and their requirements, limiting the design of the perturbation to processes that, while privatizing the results, do not jeopardize their usefulness. In this paper, we propose a privatization technique called \emph{Zeal}, where perturbator and aggregator are designed as a unit, resulting in a locally differentially private mechanism that, by-design, improves the compressibility of the perturbed dataset compared to the original, saves on transmitted bits for data collection and protects against a privacy vulnerabilities due to floating point arithmetic that affect other state-of-the-art schemes. We prove that the utility error on querying the average is invariant to the bias introduced by \emph{Zeal} in a wide range of conditions, and that under the same circumstances, \emph{Zeal} also guarantee protection against the aforementioned vulnerability. Our numerical results show up to 94~\% improvements in compression and up to 95~\% more efficient data transmissions, while keeping utility errors within 2~\%.

\end{abstract}

\section{Introduction}
The way we collect, process and store data is usually designed to achieve goals like low communication costs, private queries or reduced database size. These are generally competing objectives, and tend to be addressed with techniques that specialize in a single aspect, without synergizing with each other. For sensitive data, it is important to use privatization strategies to avoid leakage of information about their sources or owners. The design of a privatization mechanism depends on the queries it is meant to protect against, since its effectiveness for different data aggregations, like averaging or finding extrema, might vary. Queries are usually considered as inputs to the system, which limits the way we can perturb data without causing unacceptable errors on the output. Moreover, when dealing with floating point numbers, some privatization techniques relying on adding noise to the original samples are vulnerable to privacy leaks due to floating point arithmetic \cite{mironov2012significance}.
\subsection{Contribution}

This paper proposes a novel privatization and data-sharing method based on local differential privacy (LDP) called $\techniquename$, which combines the design of both the perturbation algorithm and the query by extending the \textit{piecewise mechanism} in \cite{wang2019collecting} with the \textit{addition transform} in \cite{additionmethod}. The combination of these mechanisms with a judicious selection of $\techniquename$ parameter $\Abar$, allow to achieve improving compression and transmission savings for growing $\Abar$ with low utility errors, while simultaneously protecting against floating point number vulnerabilities that would otherwise affect the original piecewise mechanism.

This vulnerability has other solutions in the literature, ranging from approximating the output of the perturbator \cite{mironov2012significance}, to strategies using smart iterative noise sampling  \cite{tinybits} or integer implementations for floating point perturbation \cite{GooglePrivacy2020}. We argue that $\techniquename$, while being equally effective in solving the vulnerability, require simpler computations and theoretical analyses.
\begin{figure}[!t]
	\centering
	\includegraphics[width =1.0 \columnwidth]{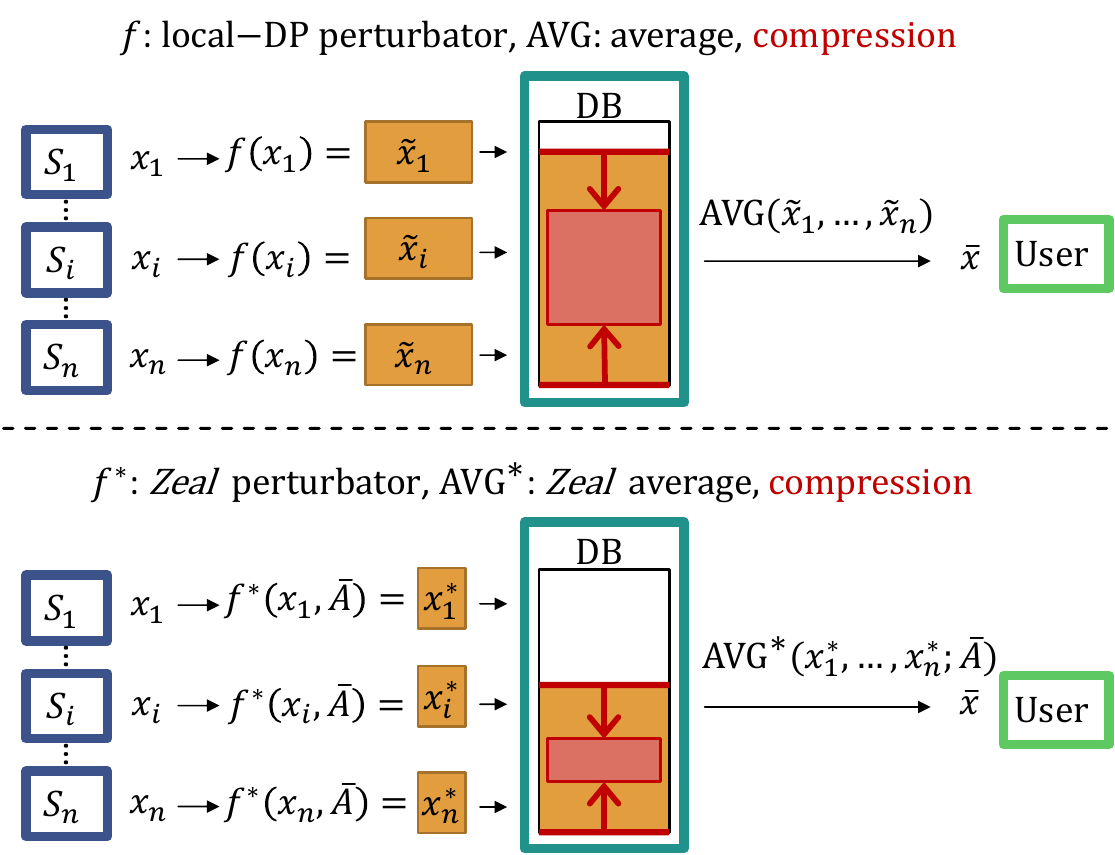}
	\caption{Comparison between a standard perturbator - aggregator scheme and $\techniquename$. In the latter, smaller packets have to be transmitted to the database, and they are more compressible than the original, while achieving the same result.}
	\label{fig:privaddIntro}
	\vspace{-0.5cm}
\end{figure}
	There are also other attempts of studying the relation between privatizing data and its compressibility. The results in \cite{papadimitriou2007time} show how specific data representations, like the wavelet transform \cite{percival2000wavelet}, when combined with proper perturbation can achieve better compressibility of data under specific privacy constraints. However, those techniques rely on privacy notions different from differential privacy, and focus on how to avoid possible vulnerabilities in the choice of additive noise. In \cite{bassily2015local}, the aim is to provide minimal private representations of the data transmitted to the database by building succinct histograms. This proves to be effective in terms of accuracy for frequency estimation, but does not cover the query of the average nor investigate the design of dedicated aggregators in combination to perturbators to enhance compression. 

\subsection{System model}
\label{systemmodel}
We consider a system illustrated in \autoref{fig:privaddIntro}, where $n$ IoT sensors of the same kind $\left\{S_1,\dots, S_n \right\}$ need to send a single floating point number $\point_i$ each to an untrusted database $\mathrm{DB}$ for storage. We assume that the final user is interested in querying the $\mathrm{DB}$ for the average of the dataset $\mathrm{DS} = \left\{\point_1,\dots,\point_n\right\}$, where the minimum and maximum value the sensors are reasonably expected to produce are represented using $\Ebar$ and $\E$ so that $\point_i \in \left[\Ebar - \E, \Ebar + \E\right]$. 
Moreover, the data being collected are sensitive, therefore mandating the need of being privatized with a perturbation function $f\left(\point_1,\dots,\point_n\right)$ before reaching the $\mathrm{DB}$. We propose a specific perturbator $\point^*_i = \techperturb\left(\point_i, \Abar\right)$, where the $\Abar$ is a publicly known parameter common to all sensors, in combination with a modified version of the average query, namely $\techaverage\left\{\point^*_1, \dots \point^*_n, \Abar\right\}$. Future work can consider couplings where each sensor may have a different $\Abar$. This new method guarantees equal privatization level, while allowing for transmission savings and better compressibility in the $\mathrm{DB}$. An example is a pharmaceutical company monitoring the average glucose measurements of their devices for patients with diabetes. 
Throughout this paper, we rely on the floating point standard IEEE-754 \cite{IEEE754} with double precision (64-bits).

\begin{table*}[!t]
	\centering
	\caption{Nomenclature}
	\vspace{-0.2cm}
		\resizebox{!}{1.8cm}{
		\begin{tabular}{rl|rl|rl}
			\toprule
			$\Abar$&\hyperref[zeal]{Bias of $\techperturb$}&\ws{ $\mathrm{CR}$}&\hyperref[eq:cr]{\w{Compression Ratio}}&\ws{ $\mathrm{TR}$}&\hyperref[eq:tr]{\w{Transmission Ratio}}\\
			
			$\point$&\hyperref[systemmodel]{$\mathrm{DS}$ sample}&\ws{$\mathrm{DS}$, $\mathrm{DS}^*$}&\hyperref[systemmodel]{\w{Dataset and privatized dataset}}&\ws{$\sumofds$}&\hyperref[thm:relativeBoundOnProbOfError]{\w{Sum of $\point_i$ in $\mathrm{DS}$}}\\
			
			\ws{$\point^*_{\max}$ }&\hyperref[zeal]{\w{Max feasible $\point_i^*$}}&\ws{$\Delta_{\mathrm{AVG}}$}&\hyperref[eq:absoluteerrorontheaverage]{\w{Absolute error on the average}}&\ws{ $\ulp(x)$}&\hyperref[eq:ulp]{\w{Unit in the last place of $x$}}\\
			
			\ws{$\point^*_{\min}$}&\hyperref[zeal]{\w{Min feasible $\point_i^*$}}&\ws{	$\techaverage$}&\hyperref[eq:techaggregator]{\w{$\techniquename$ aggregator}}&\ws{ $\elle(\mathord{\cdot})$, $\R(\mathord{\cdot})$}&\hyperref[subsec:pwmechext]{\w{Parameters of $\pdf$}}\\

			$E$&\hyperref[eq:floatingpoint]{Exponent}&LDP&\hyperref[def:ldp]{Local differential privacy}&\ws{
				$\deltaavg$}&\hyperref[eq:relativeerroraverage]{\w{Relative error on the average}}\\
			
			$\point^*$&\hyperref[systemmodel]{Output of $\techperturb$}&\ws{$\pcumul$}&\hyperref[floatingpointvulnerability]{\w{Cumulative probability}}&\ws{ $\deltaexpi$}&\hyperref[eq:relativeerrorexpected]{\w{Relative error on expected value of $\point^*_i$}}\\

			\ws{$\techperturb$}&\hyperref[subsec:pwmechext]{\w{$\techniquename$ perturbator}}&\ws{$\mathcal{F}$}&\hyperref[eq:f]{\w{Error estimation due to $\Abar$}}&\ws{	$n$}&\hyperref[systemmodel]{\w{Number of samples in $\mathrm{DS}$}}\\

			$f$&\hyperref[fig:privaddIntro]{Perturbator}&\ws{ $\E, \Ebar$ }&\hyperref[systemmodel]{\w{Parameters on feasible $\mathrm{DS}$}}&\ws{$\gamma_{\min}$}&\hyperref[eq:transmissionsaving]{\w{Min number of shared bits per sample}}\\
			
			\ws{$E_U$}&\hyperref[eq:floatingpoint]{\w{Unbiased exponent}}&\ws{$p$}&\hyperref[subsec:pwmechext]{\w{$\pdf$ probability level }}&$P(\cdot)$ &Probability\\

			\ws{ $E_U^*$}&\hyperref[thm:abarselection]{\w{Selected unbiased exponent}}&\ws{$\epsilon$ }&\hyperref[def:ldp]{\w{Privacy budget}}& &\\
			
			\bottomrule
		\end{tabular}
	}
	\vspace{-0.5cm}
	\label{tab:nomenclature}
\end{table*}

\section{Background}
\subsection{Local differential privacy}
The main goal of differential privacy is to give a bound on the likelihood that the result of a query to a dataset will allow unwanted insights on its original values \cite{dwork2006differential}. This is achieved by using a \textit{perturbator function}, where the cost for this privatization is that the private query result deviates from the one on the original data: we refer to this as \textit{utility error}. In the \textit{central model} of differential privacy \cite{dwork2006differential}, the randomization is applied directly on the collection of real data, which implies the need of a trusted party to receive the data and perturb them. The \textit{local model} assumes the presence of no trusted party, meaning that the randomization needs to occur before data is shared with the $\mathrm{DB}$.
Despite resulting in generally higher utility errors, the local version intuitively provides a more secure way of collecting and sharing data. Data leakages from the $\mathrm{DB}$, or malicious cloud providers, pose less of a threat than in the central model, since they only hold an already privatized version of the sensitive data. We exploit the fact that the local privatization needs to happens between sensors and $\mathrm{DB}$ to design a perturbator reducing transmission data size.
In order to quantify privacy, we use the \textit{privacy budget} $\epsilon > 0$ which controls the privacy-utility trade off: the more privacy we want, the  lower $\epsilon$ we should choose.
\begin{defn}\label{def:ldp}
	A perturbation function $f$ is $\epsilon$-locally differentially private if and only if for any two inputs $\point_i \neq \point_j$ in the domain of $f$, and for any output $\point^*$ of $f$, we have
	\begin{equation}
		P\left[f(\point_i) = \point^*\right] \leq e^\epsilon \cdot P\left[f(\point_j) = \point^*\right].
		\label{ldp}
	\end{equation}

\end{defn}
Given a dataset $\mathrm{DS} = \left\{\point_1, \point_2, \dots, \point_n\right\}$ and a privacy budget $\epsilon$, we want to transform each $\point_i$ to $\point^*_i$ with a randomizer $f:\point_i\rightarrow\point^*_i$ so that \autoref{ldp} holds.

\subsection{Piecewise LDP mechanism}
\begin{figure}[!t]
	\centering
	\includegraphics[width =0.8 \columnwidth]{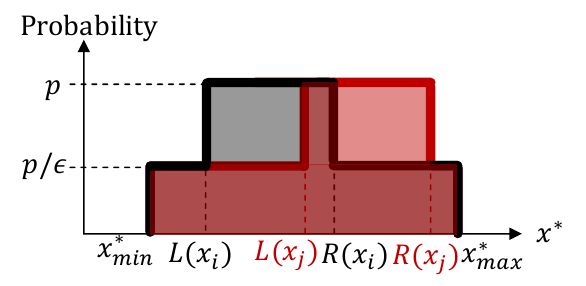}
	\vspace{-0.2cm}
	\caption{Example of piecewise mechanism $\pdf$. After perturbation, the privatized version of both $\point_i$ and $\point_j$ could be any $\point^* \in \left[\pointstarmin; \pointstarmax\right]$. }
	\label{fig:pwmechPDF}
	\vspace{-0.5cm}
\end{figure}

The original locally differentially private piecewise mechanism \cite{wang2019collecting} is a randomizer $f$ whose key features are its bounded output $\point^*_i \in [\pointstarmin, \pointstarmax]$, its lower utility error compared to other locally differentially private mechanisms, like the Laplace transformation \cite{dwork2006differential}, and the fact that it is unbiased, namely the expectation of $\point_i^*$ is $\expectedvalue[\point_i^*] = \point_i$. Another peculiar characteristic is that the probability density ($\pdf$) from which the privatized values $\point_i^*$ are sampled, changes shape according to $\point_i$, while $\pointstarmin$ and $\pointstarmax$ are fixed, as per \autoref{fig:pwmechPDF}. Wang et al. \cite{wang2019collecting} limited the original perturbator to $\mathrm{DS}$ with $\point_i \in \left[-1.0, 1.0\right]$, mentioning the steps to generalize it to $\mathrm{DS}$ in $\point_i \in \left[-r,r\right]$ by scaling the dataset. In \autoref{subsec:pwmechext} we propose a generalization of the perturbator itself so that $\point_i \in \left[\Ebar - \E, \Ebar + \E\right]$ and introduce the parameter $\Abar$ for introducing a bias to the output.
\subsection{Addition transform and floating point compression}
\label{additiontransformexplanation}
Data compression algorithms usually rely on identifying repeated symbols or sequences in the data being processed, in order to provide a compact representation of the same information. Considering a dataset of $n$ floating point numbers $\mathrm{DS} = \{\point_1,\dots, \point_n\}$, a specific kind of repeated pattern is when the $j-th$ bit of the 64 bits in each $\point_i$ has the same binary value $\forall \point_i \in \mathrm{DS}$. In this case, it is a shared bit. The addition transform\cite{additionmethod} aims to improve $\mathrm{DS}$ compressibility by increasing the number of shared bits in a dataset upon transforming the original values through the addition
\begin{equation}
	\tilde{\point}_i = \point_i + \bar{A} \quad \forall \point_i \in \mathrm{DS}.
\label{eq:additiontransform}
\end{equation}
The reason why the new $\mathrm{DS}$ typically has more shared bits than $\mathrm{DS}$ lies in the binary representation of a floating point number $x$, which is composed of a sign, an exponent and a mantissa. We refer to them as $\nu$, $E$ and $M$, when interpreted as unsigned integer, and use them to compute $x$ with
\begin{equation}
	x = (-1)^\nu \cdot 2^{E_U}\cdot(1 + M\cdot2^{-\mantissalength}),
	\label{eq:floatingpoint}
\end{equation}
where $E_U = E - b$ is the unbiased exponent, $b$ is its bias and $L$ is the length of the mantissa in bits: in double precision, $b = 1023$ and $\mantissalength = 52$. From \autoref{eq:floatingpoint}, we notice that the smallest difference between two numbers $x_1>x_2$ having the same unbiased exponent $E_U^*$ is $x_1 - x_2 = 2^{E_U^*-52}$. It is called \textit{unit in the last place}, or $\ulp(\mathord{\cdot})$, defined as
\begin{equation}
	\ulp\left(x \in \left[2^{E_U^*}, 2^{E_U^*+1}\right)\right)=2^{E_U^*-52}.
	\label{eq:ulp}
\end{equation}
$\ulp(x)$ effectively represents the \textit{precision} of $x$, since all $x^\prime \in \left(x - \ulp(x)/2; x+ \ulp(x) / 2\right)$, assuming rounding to nearest, will be represented and stored as $x$. 
If we apply \autoref{eq:additiontransform} choosing $\Abar$ large enough so that we can fit all $\tilde{\point_i}$ in a single exponent region $\left[2^{E_U^*}, 2^{E_U^*+1}\right)$, the $k$ most significant bits in $\tilde{\point_i}$ mantissas will be shared when
\begin{equation}
	\exists j \in \left[0,2^k-1\right]\subset\mathbb{N}: 
	\tilde{\point_i} \in 2^{E_U^*}\cdot\left[1 + \frac{j}{2^k}, 1 + \frac{j+1}{2^k}\right) .
\end{equation}

Note that the transformation in \autoref{eq:additiontransform} is lossy, as shown in \cite{additionmethod}. A function $g$ is lossy when, for some $\point_i$
\begin{equation}
g^{-1}\left(g\left(\point_i\right)\right) \neq \point_i.
\end{equation}

\begin{section}{Zeal}
\label{zeal}
The original formulation of the piecewise mechanism is susceptible to a privacy vulnerability due to floating point arithmetic. Since the image of the perturbator $f\left(\point_i\right)$ is a finite set of floating point numbers $\mathcal{I}_{\point_i}$, and for some $\point_i \neq \point_j$, $\mathcal{I}_{\point_i} - \mathcal{I}_{\point_j} \neq \O$, some privatized samples can be traced back to their original datum, breaking  \autoref{ldp}. $\techniquename$'s perturbator $f^*\left(\point_i, \Abar\right)$ extends $f\left(\point_i\right)$ feasible inputs and introduces the bias $\Abar$: in \autoref{floatingpointvulnerability}, we show that the extended images $\mathcal{I}^*_{\point_i}$ and $\mathcal{I}^*_{\point_j}$ guarantee  $\mathcal{I}^*_{\point_i} - \mathcal{I}^*_{\point_j} = \O$ $\forall \point_i, \point_j \in \mathrm{DS}$, solving the vulnerability. Moreover, since a similar bias is used as part of the addition transform to improve compressibility, we benefit from both advantages with the same concept.
\subsection{Extension of Piecewise Mechanism}
\label{subsec:pwmechext}
Given the input dataset $\mathrm{DS} = \left\{\point_1, \dots \point_n\right\}$ and the desired privacy level $\epsilon$, the $\pdf$ of the output $\point_i^*=\techperturb\left(\point_i, \Abar\right)$ is
\begin{equation}
	P\left(\techperturb\left(\point_i\right) = \point_i^*\right)=
	\begin{cases}
		\frac{p}{e^\epsilon}  & \text{if }  \point_i^* \in  \left[\point^*_{\min},\elle\left(\point_i\right)\right)\\
		p & \text{if } \point_i^* \in \left[\elle(\point_i), \R(\point_i)\right]\\
		\frac{p}{e^\epsilon}  & \text{if }  \point_i^* \in  \left(\R(\point_i),\point^*_{\max} \right]\\
	\end{cases}
	\label{eq:pdf}
\end{equation}
where $\point_i^* \in \left[\point^*_{\min} ,\point^*_{\max}  \right]\forall \point_i \in \mathrm{DS}^*$, with $\mathrm{DS}^*$ being the privatized dataset $\left\{\point^*_1, \dots, \point_n^*\right\}$, and
\begin{equation}
\left(\point^*_{\min}, \point^*_{\max} \right) =  \left(\Ebar - \C + \Abar, \Ebar + \C + \Abar\right),
\label{eq:outputbounds}
\end{equation}
\begin{equation}
	p = \left(e^{\epsilon}-e^{\epsilon/2}\right)/\left(2\E\left(e^{\epsilon/2} + 1\right)\right),
\end{equation}
\begin{equation}
	\C = \E \cdot\left(e^{\epsilon/2}+1\right)/\left(e^{\epsilon/2}-1\right).
	\label{eq:c}
\end{equation}
The elements that cause the $\pdf$ to vary depending on $\point_i$ are
\begin{equation}
	\elle(\point_i) = \frac{\C + \E}{2} \left(\frac{\point_i - \Ebar}{\E}\right)  - \frac{\C -\E}{2} + \Ebar + \Abar,
\end{equation}
\begin{equation}
	\R(\point_i) = \elle(\point_i) + \C - \E.
\end{equation}
$\techperturb$ is $\epsilon$-locally differentially private, since the original proof in \cite{wang2019collecting} holds. $\Abar$ introduces a bias to the output, as per \autoref{thm:privaddexpectedvalue}, while $\Ebar$ and $\E$ affect the variance according to \autoref{thm:privaddvariance}.
\begin{thm}\label{thm:privaddexpectedvalue}
Given $\point_i^* = \techperturb\left(\point_i, \Abar\right)$, its expected value is
\begin{equation}
\expectedvalue \left[\point_i^*\right] = \point_i + \Abar.
\label{eq:expectedvalue}
\end{equation}
\end{thm}
\begin{proof}
\begin{align*}
	\begin{split}
		&\expectedvalue \left[\point_i^*\right] =  \int_{\point_{\min}^*}^{\elle(\point_i)} \frac{p}{e^{\epsilon}}x \,dx +  \int_{\elle(\point_i)}^{\R(\point_i)} px \,dx + \int_{\R(\point_i)}^{\point_{\max}^*} \frac{p}{e^{\epsilon}}x \,dx  \\
		&= \frac{p\left( 1-e^\epsilon\right)}{2e^\epsilon}\left[\elle^2 - \R^2\right] + \frac{p}{2e^\epsilon}\left[4\left(\Ebar + \Abar\right)C\right] = \point_i + \Abar \\
	\end{split}
\end{align*}
\end{proof}
\begin{thm}\label{thm:privaddvariance}
Given $\point_i^* = \techperturb\left(\point_i, \Abar\right)$, its variance is
\begin{equation}
	\mathrm{VAR}(\point_i^*) = \E^2\cdot\left(\frac{(\frac{\point_i-\Ebar}{\E})^2}{e^{\epsilon/2}-1} + \frac{e^{\epsilon/2}+3}{3(e^{\epsilon/2}-1)^2}\right) .
	\label{eq:variance}
\end{equation}
\vspace{-0.3cm}
\end{thm}

Since we introduce the same bias $\Abar$ to all samples of $\mathrm{DS}^*$, we propose to remove it by using an altered version of the standard average computation, defined as
\begin{equation}
	\techaverage\left(\mathrm{DS}^*,\Abar\right) = \frac{1}{n}\sum_{i=1}^{n} \point^*_i - \Abar.
\label{eq:techaggregator}
\end{equation}

\textbf{Probability bounds of the error on the average}: The error $\Deltaavg$ between the calculated average from the privatized database and the original one is defined as
\begin{equation}
	\Deltaavg = \techaverage\left(\mathrm{DS^*}\right) - \mathrm{AVG}\left(\mathrm{DS}\right).
\label{eq:absoluteerrorontheaverage}
\end{equation}

Since $\mathrm{DS^*}$ is the output of a stochastic process with variance as per \autoref{eq:variance}, it is possible to formulate a probabilistic bound on $\left|\Deltaavg\right|$ according to \autoref{thm:boundOnProbOfError}.

\begin{thm}\label{thm:boundOnProbOfError}
Given a dataset $\mathrm{DS} = \{\point_1, \dots, \point_n\}$ and a utility error $\lambda\geq0$, an upper bound on the probability of the absolute value of the error being greater than $\lambda$ is
\begin{equation}
	P\left(\left|\Deltaavg\right|\geq \lambda\right)	\leq  e^{ - \frac{\frac{1}{2}\left(n\lambda\right)^2}{\sum_{i=1}^{n}\variance(\point^*_i) + \frac{1}{3}(\C + \E)n\lambda} }.
	\label{eq:boundOnProbOfError}
\end{equation}
\end{thm}
\begin{proof}
We can use the independent and unbiased random variable $V_i = \point_i^* -\Abar -\point_i$ to write
\begin{equation}
	P\left(\left|\Deltaavg\right|\geq \lambda\right)	= P\left(\left| \sum_{i=1}^{n}V_i \right| \geq n\cdot\lambda\right) .
	\label{eq:diffinaverage}
\end{equation}
We can reach the formulation in \autoref{eq:boundOnProbOfError} using the Bernstein inequality together with the output bounds in \autoref{eq:outputbounds}, the sampling expected value in \autoref{eq:expectedvalue} and variance \autoref{eq:variance}.
\end{proof}

Since we are more interested in expressing the error on the average relatively to the original average, which is
\begin{equation}
	\deltaavg = \left(\techaverage\left(\mathrm{DS^*}\right) - \mathrm{AVG}\left(\mathrm{DS}\right)\right)/\mathrm{AVG}\left(\mathrm{DS}\right),
\label{eq:relativeerroraverage}
\end{equation}
we can adapt the bound in \autoref{thm:boundOnProbOfError} to have a relative formulation using $\deltaavg$, as per \autoref{thm:relativeBoundOnProbOfError}.

\begin{thm}\label{thm:relativeBoundOnProbOfError}
Given a dataset $\mathrm{DS} = \{\point_1,\dots, \point_n\}$, $\mathrm{S}_{\mathrm{DS}} = \sum_{i=1}^{n} \point_i$ and a utility error $\lambda\geq0$, an upper bound on the probability of the absolute value of the relative error being greater than $\lambda$ is
	\begin{equation}
		P\left(\left|\deltaavg \right|\geq \lambda\right) \leq e^{ - \frac{\frac{1}{2}\left(n\sumofds \lambda\right)^2}{\sum_{i=1}^{n}\variance(t^*_i) + \frac{1}{3}(\C + E)n\sumofds\lambda} }.
		\label{eq:boundOnRelativeProbOfError}
	\end{equation}
\end{thm}

\begin{remark}\label{lemma:abarinvariance}
Since neither \autoref{eq:boundOnProbOfError} nor \autoref{eq:boundOnRelativeProbOfError} depend on $\Abar$, both error bounds are $\Abar$ invariant. In \autoref{results} we empirically show that errors are $\Abar$ invariant as well, meaning that selecting different values of $\Abar$ result in similar $\left|\Deltaavg\right|$ and similar $\left|\deltaavg\right|$. However, this is true only when assuming infinitely precise numbers. In \autoref{Abar} we detail the reasons and show possible consequences of finite precision.
\end{remark}

\subsection{Selection of addition transform parameter}
\label{Abar}
The $\Abar$ parameter allows us to manipulate the binary representation of the private dataset so that some bits are shared by all $\point_i^*$.
An effective selection of $\Abar$ fulfilling the recommendations in \cite{additionmethod} and the conditions in \autoref{additiontransformexplanation}, while guaranteeing that all sign and exponent bits are shared, is presented in \autoref{thm:abarselection}.
\begin{thm}\label{thm:abarselection}
	Given a dataset $\mathrm{DS}$, we define the unbiased exponent $\minexpencl$ of the smallest region of numbers with equal exponents that can enclose the privatized dataset $\mathrm{DS^*}$ as
	\begin{equation}
	\minexpencl = \left\lceil \log_2(2\C) \right\rceil.
	\end{equation}
	By selecting $E^*_U \geq \minexpencl >-1022\in\mathbb{Z}$,
	we compute $\Abar$ as
	\begin{equation}
		\Abar =  2^{{E^*_U}+1} - 2\cdot \ulp\left(2^{E^*_U}\right) - \Ebar - \C.
		\label{eq:alignds}
	\end{equation}
	This formulation ensures that all sign and exponent bits are shared $\forall \point_i^* \in \mathrm{DS^*}$. The larger the selected $E^*_U$, the more mantissa bits will be shared as well.
\end{thm}

The selection of $\Abar$ should also take into consideration the loss due to the floating point finite precision, since the expected value in \autoref{eq:expectedvalue}, and the error bounds in \autoref{eq:boundOnProbOfError} and \autoref{eq:boundOnRelativeProbOfError} hold only assuming infinite precision. However, the error with finitely precise numbers is significant only when $\Abar$ is computed according to \autoref{eq:alignds} with ${E^*_U} \gg \minexpencl$, potentially resulting in unacceptable alterations to $\techniquename$.

\begin{figure}[!bt]
	\centering
	\includegraphics[width =1.0 \columnwidth]{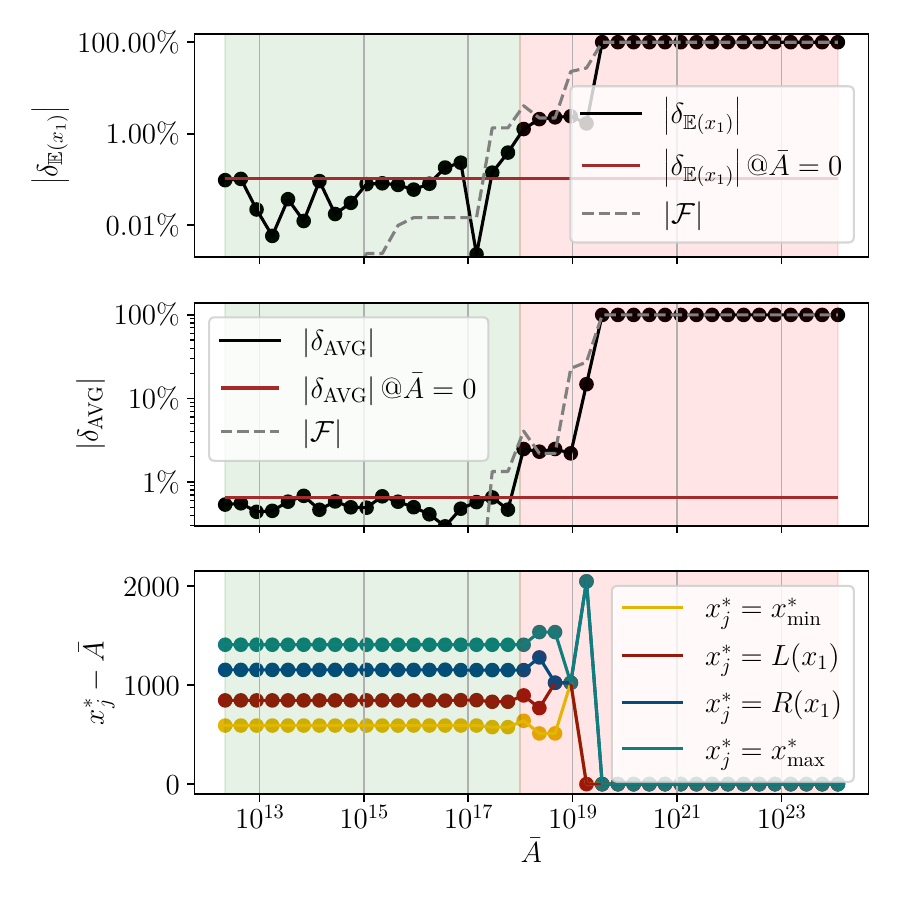}
	\vspace{-1.0cm}
	\caption{
		Effects of large $\Abar$ on expected value, $\deltaavg$ and $\pdf$, with $\mathcal{F}$ as an estimation. $\point_j^*$ in the bottom plot are unbiased to be able to compare them.}
	\vspace{-0.5cm}
	\label{fig:AbarTooBigError}
\end{figure}

We examine the effects of large $\Abar$ on $\techniquename$ by considering a uniformly distributed $\mathrm{DS}$ with $n = 1000$, $\Ebar = 1000$, $\E = 100$ and the privatization $\techperturb\left(\mathrm{DS}, \Abar\right)$ for various values of $\Abar$, as per \autoref{fig:AbarTooBigError}. We first investigate the expected value of $\point_1^* = \techperturb(\point_1, \Abar)$ with $\point_1 = 1000$ by plotting $\deltaexpone$, defined as
\begin{equation}
	\deltaexpi = \left(\mathbb{E}(\point_i^*) - \Abar - \point_i\right) / \point_i,
\label{eq:relativeerrorexpected}
\end{equation}
where $\mathbb{E}(\point_1^*)$ is estimated by averaging $10^5$ samples of $\point_1^*$. For this dataset, when $\Abar \geq 10^{18}$ (red area), $\deltaexpone$ starts to deviate from the ideal $0\%$ we would have with infinite precision and infinite samples, clipping to $100\%$ for $\Abar \geq 10^{20}$ since
\begin{equation}
\techperturb(\point_1) = \Abar \implies \mathbb{E}(\point_1^*) = \Abar.
\label{eq:destructiveAbar}
\end{equation} 
When \autoref{eq:destructiveAbar} is true ${\forall \point_i \in \mathrm{DS}}$, there is
\begin{equation}
	\techaverage\left(\point_1^*, \dots , \point_n^*\right) = 0.0, 
\end{equation}
which causes the  $\left|\deltaavg\right|$ to clip at $100\%$ as well. It should be noted that $\left|\deltaavg\right|$ could potentially be greater than $100\%$ for some $\Abar$, since this approximation effects drastically change $\techperturb$ $\pdf$ as well, as we can see in the plot at the bottom of \autoref{fig:AbarTooBigError}. For $\Abar \geq 10^{20}$, $\left(\pointstarmin, \elle(\point_1), \R(\point_1), \pointstarmax\right) - \Abar= \left(0, 0, 0, 0\right)$.

In order to estimate both $\deltaexpone$ and $\deltaavg$ for large $\Abar$ without knowing $\mathrm{DS}$ but only $\E$ and $\Ebar$ we introduce $\mathcal{F}$, computed as the average of the approximation error on the values $\Ebar - \C$ and $\Ebar + \C$ due to $\Abar$. $\mathcal{F}$ is defined as
\begin{equation}
	\mathcal{F} = \left[\left(\Delta\pointstarmin/\left(\Ebar - \C\right)\right) + \left(\Delta\pointstarmax/\left(\Ebar + \C\right)\right)\right]/2,
\label{eq:f}
\end{equation}
where $\Delta\pointstarmin = \Ebar - \C  - \left(\pointstarmin - \Abar\right)$ and $\Delta\pointstarmax = \Ebar + \C  - \left(\pointstarmax - \Abar\right)$. In \autoref{fig:AbarTooBigError} we plot $\mathcal{F}$ and notice that it is pretty accurate in describing when $\Abar$ is too large to maintain acceptable error. Therefore, we can use it to select an appropriate $\Abar$.

\subsection{Transmission savings}
\label{transmissionsavings}

As discussed in \autoref{Abar}, the selection of $\Abar$ depends only on $\Ebar$, $\E$ and $\epsilon$, thus $\Abar$ can be fixed even before the sensors are deployed. Assuming that $\Abar$ is chosen as per \autoref{eq:alignds}, we can guarantee that a minimum number of bits per sample $\gamma_{\min}$, computed according to  \autoref{thm:guaranteedcommonbits}, is shared by all $\point^*_i$ and whose binary value is known a priori. Therefore, if each one of the $n$ sensors transmits only $64 - \gamma_{\min}$ bits per sample, the database will still be able to reconstruct $\mathrm{DS^*}$. We measure the savings in communication costs with this strategy by defining the \textit{transmission ratio} ($\mathrm{TR}$) as
\begin{equation}
	\mathrm{TR} \left(\mathrm{DS}\right) = \frac{\mathrm{DS}\text{ size in bits} - n\cdot\gamma_{\min}}{\mathrm{DS}\text{ size in bits}} = 1-\frac{\gamma}{64}.
\label{eq:tr}
\end{equation}

\begin{thm}\label{thm:guaranteedcommonbits}
	Given a privatized dataset $\mathrm{DS^*} = \techperturb\left(\mathrm{DS}, \Abar\right)$ with $\Abar$ computed according to \autoref{eq:alignds}, the minimum number of guaranteed bits shared per sample in $\mathrm{DS^*}$ is
\begin{equation}
	\gamma_{\min} = 1 + 11 + E_U^*- \left\lceil\log_2(2\cdot\C + 3\cdot\ulp\left(2^{E_U^*}\right)\right\rceil.
	\label{eq:transmissionsaving}
\end{equation}
\end{thm}
\begin{proof}
	Computing $\Abar$ as per \autoref{eq:alignds} ensures that
	\begin{equation}
	\point^*_i \in \left[2^{{E_U^*}+1} - 2\cdot\C - 3\cdot\ulp\left(2^{E_U^*}\right), 2^{{E_U^*}+1}\right)\, \forall \point_i^*.
	\end{equation}
	To represent all floating point numbers in the interval, we need the number $m$ of changing mantissa bits to be such that
	\begin{equation}
		2^m \cdot \ulp\left(2^{E_U^*}\right) \geq 2\cdot\C + 3\cdot\ulp\left(2^{E_U^*}\right), 
	\end{equation} 
	from which we can find $m_{\min}$. The expression in \autoref{thm:guaranteedcommonbits} follows by summing the count of guaranteed shared mantissa bits, namely $52 - m_{\min}$, with 1 for the single sign bit and 11 for the exponent bits, that are all shared due to \autoref{eq:alignds}.
\end{proof}


\end{section}
\begin{section}{Floating point vulnerability}
\label{floatingpointvulnerability}
In differential privacy for numerical datasets, perturbing data generally involves sampling a random variable with specific $\pdf$. When the datatype is floating point, this step should be performed with particular caution, since its arithmetic rules might lead to privacy leaks, as discussed in \cite{mironov2012significance} and \cite{tinybits}. 

\begin{figure*}
	\centering
	\includegraphics[width =0.85\textwidth]{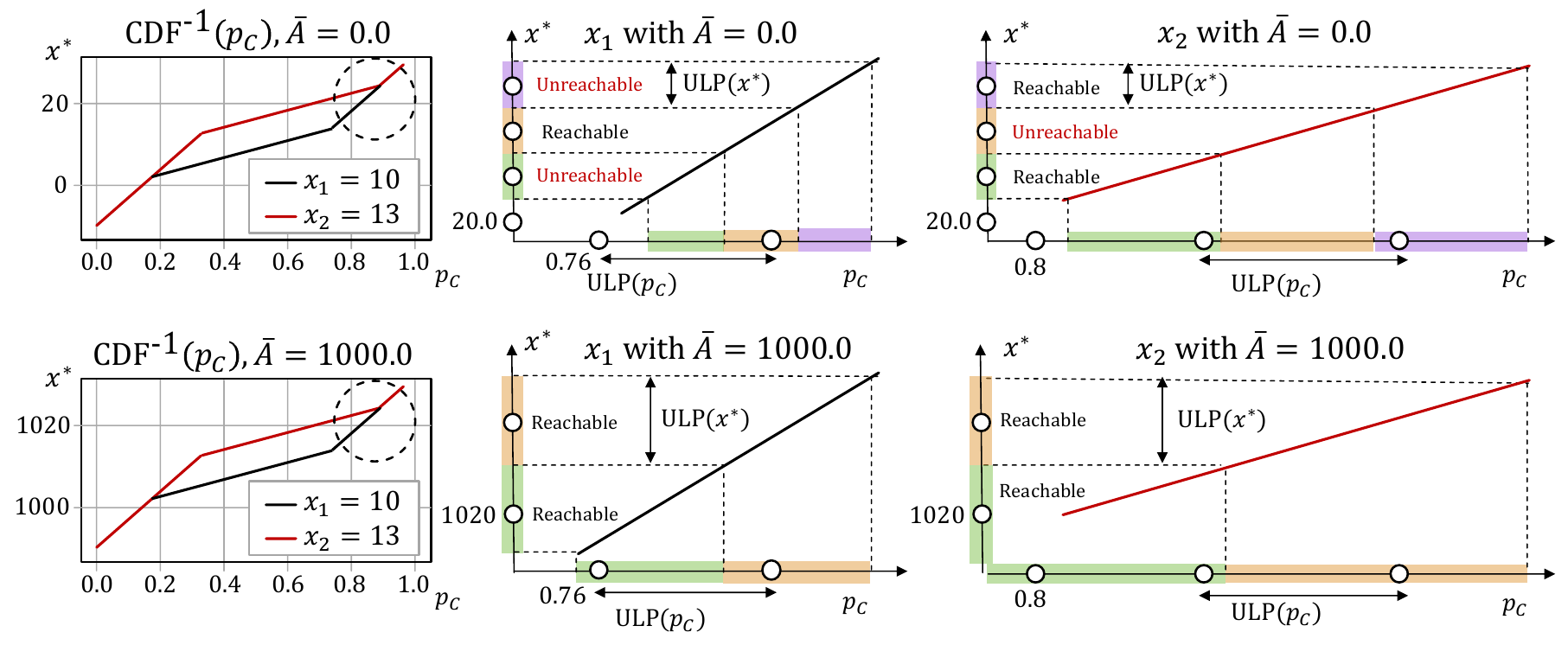}
	\vspace{-0.4cm}
	\caption{
		Floating point vulnerability for a $\mathrm{DS}$ with $\Ebar = 10$, $\E = 5$ and $\epsilon = 1$, with and without $\Abar$. When $\Abar = 0.0$, the reachable $\point^*_i$ could change depending on  $\point_i$. With a large enough $\Abar$, all $\point^*$ are reachable, causing no differentiating results that would lead to a privacy leak.
	}
	\vspace{-0.5cm}
	\label{fig:holesInvCdf}
\end{figure*}

 One of the most common methods to sample a $\pdf$ is to use its $\invcdf$, namely the inverse of its cumulative distribution function. $\invcdf$ has the cumulative probability as input, with values $p_{C} \in \left[0.0, 1.0\right)$, and $\point^*_i \in \left[ \pointstarmin, \pointstarmax\right]$ as output: it can be proved that by sampling a uniformly distributed random variable from $0.0$ to $1.0$, and then applying the $\invcdf$ on the sample, the output is distributed as the desired $\pdf$ \cite{larson1981urban}. Given $\point_i$, the vulnerability arises from the fact that the set of possible $\point^*_i$ if finite, meaning that some floating point number in $\left[\pointstarmin, \pointstarmax\right]$ might be unreachable output of $\techperturb\left(\point_i, \Abar\right)$. As depicted in \autoref{fig:holesInvCdf}, if the reachability of any $\point^*_i \in \left[\pointstarmin, \pointstarmax\right]$ depends on $\point_i$, the mechanism is not locally differentially private as per \autoref{def:ldp}.

$\techniquename$ solves the vulnerability by guaranteeing no unreachable $\point^*_i$, since the selection of $\Abar$ according to \autoref{thm:minexpvul} implies that each $\point^*_i$ is the output of at least one $\pcumul$, as per the second row in \autoref{fig:holesInvCdf}.

\begin{thm}\label{thm:minexpvul}
	$\Abar$ preventing the vulnerability described in \autoref{floatingpointvulnerability} from causing privacy leaks are computed according to \autoref{eq:alignds} with $E^*_U \geq \minexpvul$, where

	\begin{equation}
		\minexpvul = \max\left(\left\lceil -1 + \log_2 \left(e^\epsilon/p\right)\right\rceil,	\minexpencl \right).
	\label{eq:minexpvul}
	\end{equation}
\end{thm}
\begin{proof}
In order to ensure that every $\point^*_i \in \left[\pointstarmin, \pointstarmax\right]$ has a corresponding $\pcumul$ pointing to it via $\invcdf$, any interval $\invcdf(\ulp(\pcumul))$ wide should always contain at least one floating point $\point^*_i$. Since  $\invcdf$ is composed of lines with angular coefficient $\angularcoeff_i \in \{\frac{e^\epsilon}{p}, \frac{1}{p}\}$, we need
\begin{equation}
	\ulp\left(\pcumul\right) \cdot \angularcoeff_i \leq \ulp\left(\point^*_i\right),
	\label{eq:conditionforholes}
\end{equation}
where $\angularcoeff_i = \frac{e^\epsilon}{p}$ and $\pcumul  \in \left[0.5, 1.0\right)$ are the worst scenario, since steeper lines and bigger $\ulp(\pcumul)$ lead to more $\point^*_i$ being unreachable. Under these conditions, \autoref{eq:conditionforholes} becomes
\begin{equation}
	2^{E_U^* -52}\cdot 2^{53} \geq \left(e^\epsilon/p\right) \implies E_U^* \geq -1 + \log_2\left(e^\epsilon/p\right).
\label{eq:minexpvulproof}
\end{equation} 
The \autoref{eq:minexpvul} comes from combining \autoref{eq:minexpvulproof} with $E_U^* \geq \minexpencl$, since we need $\ulp\left(\point_i^*\right)$ to be constant $\forall \point_i^* \in \mathrm{DS^*}$.
\end{proof}

\end{section}
\begin{section}{Results}
\label{results}
In this section we present the results of $\techniquename$ on the first 5000 elements of the dataset \textit{``aarhus-citylab-humidity"} \cite{AarhusKommune_2017} and the first 1000 of \textit{``chicago-taxi-trips-fares"} \cite{TaxiDataset}. Given their extrema, we assume that the former has feasible values $\point_i \in [23.5, 83.9]$,
and the latter $\point_i \in [1.0, 120.0]$.
To measure compression, we use the \textit{compression ratio} ($\mathrm{CR}$), defined as 
\begin{equation}
	\mathrm{CR}\left(\mathrm{DS}\right) = \frac{\text{Compressed } \mathrm{DS}\text{ size in bits}}{\text{Uncompressed } \mathrm{DS}\text{ size in bits}}.
\label{eq:cr}
\end{equation}
The compressor used for these analyses is \textit{Greedy-GD}\cite{GD_Greedy}, which is based on bits deduplication. 
As reported in \cite{additionmethod}, \textit{Greedy-GD} benefits from an increased number of shared bits.

\begin{figure}[!bt]
	\centering
	\includegraphics[width =1.0 \columnwidth]{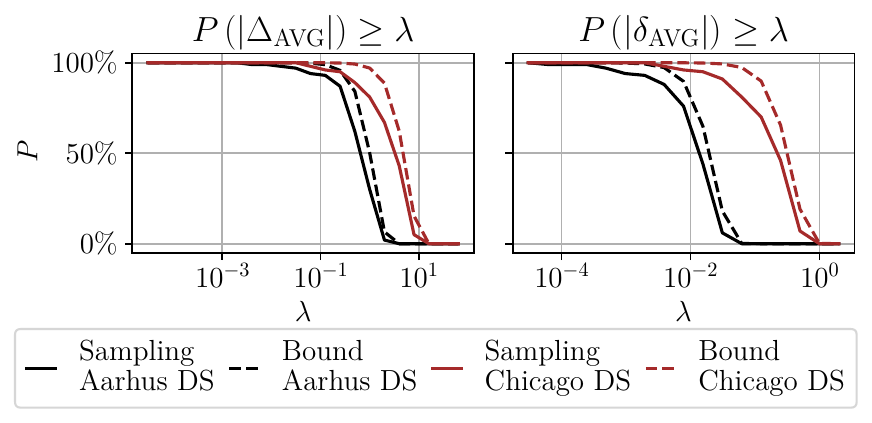}
	\vspace{-0.8cm}
	\caption{
		Probability bound on the error of the average compared to sampling.	}
	\vspace{-0.4cm}
	\label{fig:probBoundOnErrorAVG}
\end{figure}

\begin{figure}[!bt]
	\centering
	\includegraphics[width =0.9 \columnwidth]{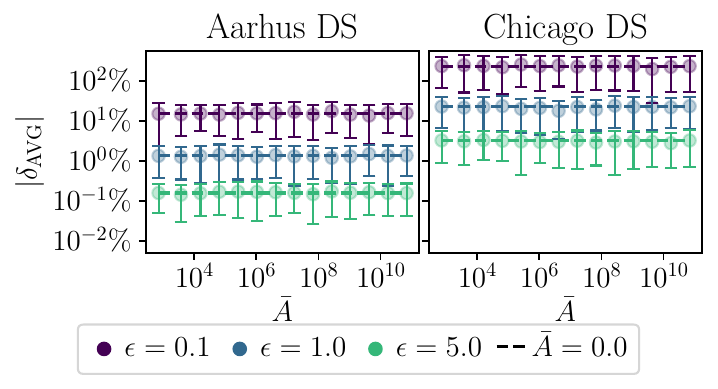}
	\vspace{-0.5cm}
	\caption{
		Relative error on the average for varying $\epsilon$ and $\Abar$.
	}
	\vspace{-0.3cm}
	\label{fig:percErrorOnAvgVaryingEps}
\end{figure}
\textbf{Utility error on the average:} In \autoref{fig:probBoundOnErrorAVG} we compare the probabilistic upper bounds on the utility error, both in absolute ($\Deltaavg$) and in relative terms ($\deltaavg$), with the probability based on samples averaged over $10$ iterations of $\techniquename$. The two are comparable, and both show that the probability decreases as $\lambda$ increases. Moreover, for sufficiently large $\lambda$ values both probabilities are guaranteed to be zero: under $\epsilon$-local differential privacy, this can not be achieved by perturbators with unbounded outputs, like the Laplace mechanism.

In order to estimate the impact of $\epsilon$  and $\Abar$ on $\techniquename$'s utility error, we compute $\deltaavg$ over $100$ iterations for multiple $\epsilon$ and $\Abar$, as shown in \autoref{fig:percErrorOnAvgVaryingEps}. We see that with smaller $\epsilon$ the perturbator has to introduce more noise to privatize the data, increasing the utility error, and that similar $\epsilon$ result in different errors depending on the dataset. Moreover, as per \autoref{lemma:abarinvariance}, different $\Abar$ values result in similar utility errors.

\begin{figure}[!bt]
	\centering
	\includegraphics[width =1.0 \columnwidth]{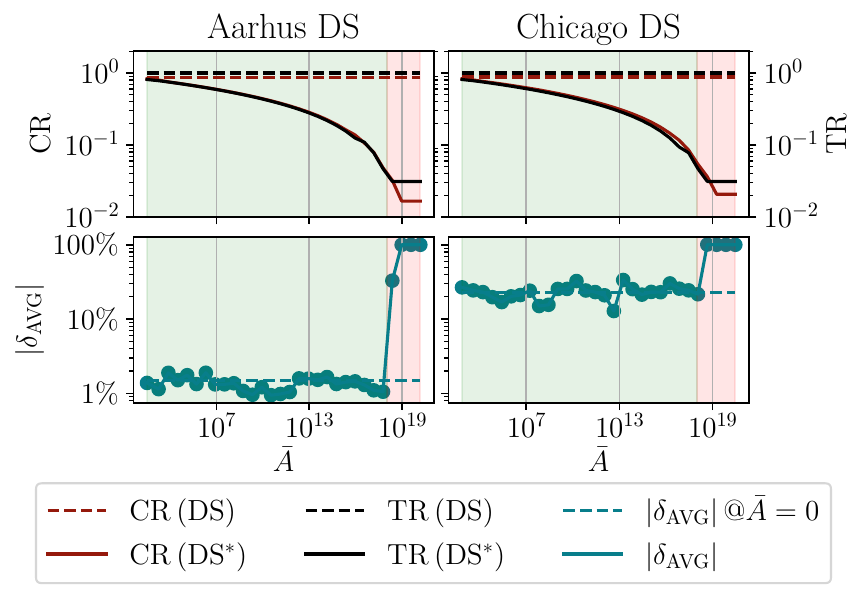}
	\vspace{-0.8cm}
	\caption{
		Transmission savings and compressibility against $\deltaavg$. Any $\Abar$ in the green area achieves improvements with comparable error levels.
	}
	\vspace{-0.9cm}
	\label{fig:CRandTR}
\end{figure}
\textbf{Transmission savings and compressibility:} In \autoref{fig:CRandTR}, we analyze transmission ratio ($\mathrm{TR}$) and compression ratio ($\mathrm{CR}$) against $\deltaavg$ for a range of $\Abar$, with $\epsilon = 1.0$. $\mathrm{TR}$ and $\mathrm{CR}$ improve with very similar rates by increasing $\Abar$, since lower $\mathrm{TR}$ means less data to transmit from the sensors to the $\mathrm{DB}$, and lower $\mathrm{CR}$ means better compressed privatized dataset. For $\Abar < 10^{18}$ (end of green area), the relative utility error $\left|\deltaavg\right|$  remains nearly the same as for $\Abar = 0$. Using the largest feasible $\Abar$ in the green area, 
we reach up to a $94\%$ improvement in $\mathrm{CR}$, and up to a $95\%$ improvement in $\mathrm{TR}$.

\end{section}
\begin{section}{Conclusions and future work}
	In this paper, we introduced $\techniquename$, a privatization technique composed of a perturbator and an aggregator for measuring the average of a dataset, which aims at reducing transmission data size and data storage, as well as solving a privacy vulnerability related to floating point data. We presented its characteristics, proved the necessary conditions for solving the vulnerability and showed its capabilities on a real-life dataset. We plan to improve $\techniquename$ so to process datasets with multiple dimensions, both in terms of attributes and as time-series, and expand the core ideas to other privatization mechanisms based on local differential privacy and to metrics other than the average.

\end{section}
\bibliographystyle{IEEEtran}
\bibliography{IEEEabrv,library,IEEEtran_control}

\begin{thebibliography}{10}
\providecommand{\url}[1]{#1}
\csname url@samestyle\endcsname
\providecommand{\newblock}{\relax}
\providecommand{\bibinfo}[2]{#2}
\providecommand{\BIBentrySTDinterwordspacing}{\spaceskip=0pt\relax}
\providecommand{\BIBentryALTinterwordstretchfactor}{4}
\providecommand{\BIBentryALTinterwordspacing}{\spaceskip=\fontdimen2\font plus
\BIBentryALTinterwordstretchfactor\fontdimen3\font minus
  \fontdimen4\font\relax}
\providecommand{\BIBforeignlanguage}[2]{{%
\expandafter\ifx\csname l@#1\endcsname\relax
\typeout{** WARNING: IEEEtran.bst: No hyphenation pattern has been}%
\typeout{** loaded for the language `#1'. Using the pattern for}%
\typeout{** the default language instead.}%
\else
\language=\csname l@#1\endcsname
\fi
#2}}
\providecommand{\BIBdecl}{\relax}
\BIBdecl

\bibitem{mironov2012significance}
\BIBentryALTinterwordspacing
I.~Mironov, ``On significance of the least significant bits for differential
  privacy,'' in \emph{CCS}, 2012. [Online]. Available:
  \url{tinyurl.com/5n7r5jsc}
\BIBentrySTDinterwordspacing

\bibitem{wang2019collecting}
\BIBentryALTinterwordspacing
N.~Wang \emph{et~al.}, ``Collecting and analyzing multidimensional data with
  local differential privacy,'' in \emph{IEEE ICDE}, 2019. [Online]. Available:
  \url{tinyurl.com/bddsmued}
\BIBentrySTDinterwordspacing

\bibitem{additionmethod}
\BIBentryALTinterwordspacing
F.~Taurone \emph{et~al.}, ``Change a bit to save bytes: Compression for
  floating point time-series data,'' in \emph{IEEE ICC}, 2023. [Online].
  Available: \url{arxiv.org/abs/2303.04478}
\BIBentrySTDinterwordspacing

\bibitem{tinybits}
\BIBentryALTinterwordspacing
S.~Haney \emph{et~al.}, ``Precision-based attacks and interval refining: how to
  break, then fix, differential privacy on finite computers,'' \emph{ICML},
  2022. [Online]. Available: \url{arxiv.org/abs/2207.13793}
\BIBentrySTDinterwordspacing

\bibitem{GooglePrivacy2020}
Google, ``Secure noise generation,'' \url{tinyurl.com/3z6fdn69}, 2020.

\bibitem{papadimitriou2007time}
S.~Papadimitriou \emph{et~al.}, ``Time series compressibility and privacy,'' in
  \emph{VLDB 2007}.

\bibitem{percival2000wavelet}
D.~B. Percival \emph{et~al.}, \emph{Wavelet methods for time series
  analysis}.\hskip 1em plus 0.5em minus 0.4em\relax Cambridge university press,
  2000, vol.~4.

\bibitem{bassily2015local}
R.~Bassily \emph{et~al.}, ``Local, private, efficient protocols for succinct
  histograms,'' in \emph{ACM STOC}, 2015.

\bibitem{IEEE754}
\emph{IEEE 754-2019 Standard for Floating-Point Arithmetic}, Std., 2019.

\bibitem{dwork2006differential}
C.~Dwork \emph{et~al.}, ``The algorithmic foundations of differential
  privacy,'' \emph{Foundations and Trends{\textregistered} in Theoretical
  Computer Science}, 2014.

\bibitem{larson1981urban}
R.~C. Larson \emph{et~al.}, \emph{Urban operations research}, 1981, ch. 7.1.3.

\bibitem{AarhusKommune_2017}
{Aarhus Kommune}, ``Sensordata,'' \url{tinyurl.com/heeth2fd}, 2017.

\bibitem{TaxiDataset}
{City of Chicago}, ``Taxi trips dataset,'' \url {tinyurl.com/4rypurjp}.

\bibitem{GD_Greedy}
\BIBentryALTinterwordspacing
A.~Hurst \emph{et~al.}, ``Greedy{GD} : Enhanced generalized deduplication for
  direct analytics in {IoT},'' 2023. [Online]. Available:
  \url{arxiv.org/abs/2304.07240}
\BIBentrySTDinterwordspacing

\end{thebibliography}

\end{document}